\documentstyle[12pt]{article}
\input epsf.sty
\input psfig.sty
\def\ga{\mathrel{\raise.3ex\hbox{$>$\kern-.75em\lower1ex\hbox{$\sim$}}}}
\def\la{\mathrel{\raise.3ex\hbox{$<$\kern-.75em\lower1ex\hbox{$\sim$}}}}
\def\he#1{\hbox{${}^{#1}$He}}
\def\li#1{\hbox{${}^{#1}$Li}}
\def\beq{\begin{equation}}
\def\eeq{\end{equation}}

\begin{document}
\pagestyle{plain}
\baselineskip=13pt
\rightline{UMN--TH--1510-11/96}
\rightline{astro-ph/9609071}
\rightline{September 1996}
\vspace*{3.2cm}
\begin{center}
{\large{
Primordial Nucleosynthesis 
\footnote{Summary of talks presented at the VIIIth 
Recontres de Blois, Neutrinos, Dark Matter, and the Universe,
Blois, France, June 8-12, 1996 and the XVII International Conference
on Neutrino Physics and Astrophysics, $\nu$'96, Helsinki,
Findland, June 13-19, 1996.}
 }}
\end{center}
~\newline

\baselineskip=2ex
\begin{center}
{\large Keith A.~Olive
}\\
{\large \it
{School of Physics and Astronomy,
University of Minnesota,\\ Minneapolis, MN 55455, USA}}

\vspace*{3cm}
{\bf Abstract}
\end{center}
The current of status of big bang nucleosynthesis is reviewed and
the concordance between theory and observation is examined in
detail.  It is argued that when using the observational data 
on \he4 and \li7, the two isotopes whose abundances are least 
affected by chemical and stellar evolution, both are completely
consistent with BBN theory. In addition, these isotopes determine
the value of the baryon-to-photon ratio, $\eta$ to be relatively low,
$\eta \approx 1.8 \times 10^{-10}$, which happens to agree with 
some recent measurements of D/H in quasar absorption
systems.  These results have far reaching consequences for
galactic chemical evolution, the amount of baryonic dark
matter in the Universe and on the allowed number of degrees of freedom
in the early Universe.  
\newpage
\vspace*{4.5ex}
\baselineskip=3ex
 
The concordance between big bang nucleosynthesis (BBN) theory and observation
has been the subject of considerable recent debate.  It is clear however,
that the real questions lie not with the concordance between BBN and 
the observational data, but rather between the theories of chemical
and stellar evolution and the data.  BBN theory (see for example \cite{wssok})
is quite stable in the sense that over time very little
in the fundamental theory
has changed.  Cross-sections are now somewhat more accurately
measured, the neutron mean life is known with a much higher degree of
precision, and if we restrict our attention to the standard model, the number
of neutrinos has also been determined. In contrast, the status of the 
observational data has changed significantly in the last several years.
There is better data on \he4, more data on \li7, and data on D and \he3 
that was simply non-existent several years ago.  For the most part, the
inferred abundances of \he4 and \li7 have remained relatively fixed, giving us 
a higher degree of confidence in the assumed primordial abundances of these
isotopes as is reflected in their observational uncertainties. Indeed, the
abundances of \he4 and \li7 alone are sufficient in order to probe and 
test the theory and determine the single remaining parameter in the standard
model, namely, the baryon-to-photon ratio, $\eta$ \cite{fo}.
In contrast, D and \he3 are highly dependent on models of chemical evolution
(\he3 is in addition dependent on the uncertain stellar yields of this isotope).
New data from quasar absorption systems, on what may be primordial D/H is at
this time disconcordant, different measurements give different abundances.
As a consequence of the uncertainties in D and \he3, 
one can use the predictions based on \he4 and \li7 in order
to construct models of galactic chemical evolution.  These results also have 
important implications for the amount of (non)-baryonic dark matter in the
galaxy and on the number of allowed relativistic degrees of freedom at the
time of BBN, commonly parameterized as $N_\nu$.

\bigskip

Before commencing with the direct comparison between theory and observations,
it will be useful to briefly review the main events leading to the synthesis
of the light elements.
Conditions for the synthesis of the light elements were attained in the
early Universe at temperatures  $T \la $ 1 MeV.  
At somewhat higher temperatures, weak interaction rates were
in equilibrium, thus fixing the ratio of number densities of neutrons to
protons. At $T \gg 1$ MeV, $(n/p) \simeq 1$.  As the temperature fell and
approached the point where the weak interaction rates were no longer fast enough
to maintain equilibrium, the neutron to proton ratio was given approximately by
the Boltzmann factor, $(n/p) \simeq e^{-\Delta m/T}$, where $\Delta m$
is the neutron-proton mass difference. The final abundance of \he4 is very
sensitive to the $(n/p)$ ratio.

The nucleosynthesis chain begins with the formation of deuterium
through the process, $p+n \rightarrow$ D $+ \gamma$.
However, because the large number of photons relative to nucleons,
$\eta^{-1} = n_\gamma/n_B \sim 10^{10}$, deuterium production is delayed past
the point where the temperature has fallen below the deuterium binding energy,
$E_B = 2.2$ MeV (the average photon energy in a blackbody is
${\bar E}_\gamma \simeq 2.7 T$).  
When the quantity $\eta^{-1} {\rm exp}(-E_B/T) \sim 1$ the rate for 
deuterium destruction (D $+ \gamma \rightarrow p + n$)
finally falls below the deuterium production rate and
the nuclear chain begins at a temperature $T \sim 0.1 MeV$.

The dominant product of big bang nucleosynthesis is \he4 resulting in an
abundance of close to 25\% by mass. This quantity is easily estimated by
counting the number of neutrons present when nucleosynthesis begins.
When the weak interaction rates freeze-out, at $T \approx 0.8$ MeV,
the neutron to proton ratio is about 1/6. When free neutron decays 
are taken into account prior deuterium formation, the ratio drops to
$(n/p) \approx 1/7$. Then simple counting yields a \he4  mass fraction
\beq
Y_p = {2(n/p) \over \left[ 1 + (n/p) \right]} \approx 0.25
\label{ynp}
\eeq

In the standard model,
 the \he4 mass fraction
depends primarily on the baryon to photon ratio,
$\eta$ as it is this quantity which determines the onset of nucleosynthesis 
via deuterium production. But because the $(n/p)$ ratio is only
weakly dependent on $\eta$, the \he4 mass fraction is relatively
flat as a function of $\eta$. When we go beyond the standard model, the
\he4 abundance is very sensitive to changes in the expansion rate which 
can be related to the effective number of neutrino flavors as will
be discussed below. Lesser amounts of the other light elements are produced:
D and \he3 at the level of about $10^{-5}$ by number, and \li7 at the level of
$10^{-10}$ by number. 

\begin{figure}[htbp]
\hspace{0.5truecm}
\epsfysize=6.8truein\epsfbox{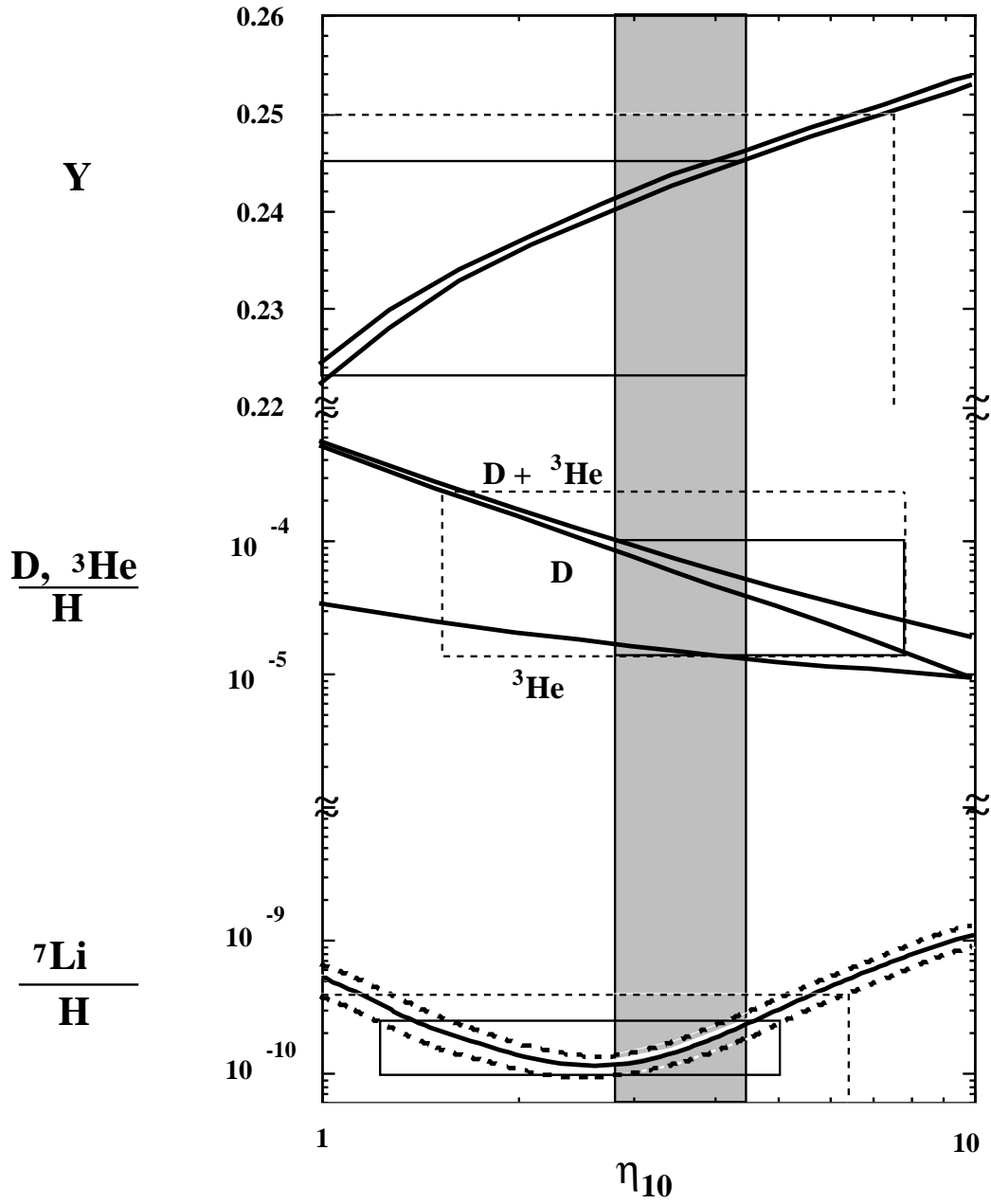}
\caption{{The light element abundances from big bang
nucleosynthesis.}}
\end{figure}

The resulting abundances of the light elements are shown in Figure 1. 
The curves for the \he4 mass fraction, $Y$, bracket the
computed range based on the uncertainty of the neutron mean-life which 
has been taken as \cite{rpp} $\tau_n = 887 \pm 2$ s. 
 Uncertainties in the produced \li7 
abundances have been adopted from the results in Hata et al.
\cite{hata1}. Uncertainties in D and
\he3 production are negligible on the scale of this figure. 
The  boxes correspond
to the observed abundances and will be discussed below. 

\bigskip

There is now a good collection of abundance information on the \he4 mass
fraction, $Y$, O/H, and N/H in over 70 extragalactic HII 
(ionized hydrogen) regions
\cite{p,evan,iz}. The observation of the heavy elements is 
important as the helium
mass fraction observed in these HII regions has been augmented by some stellar
processing, the degree to which is given by the oxygen and nitrogen abundances.
In an extensive study based on the data in \cite{p,evan}, it was found
\cite{osa}   that the data is well represented by a linear correlation for
$Y$ vs. O/H and Y vs. N/H. It is then expected that the primordial abundance
of \he4 can be determined from the intercept of that relation.  
The overall result
of that analysis indicated a primordial mass fraction, 
 $Y_p  = 0.232 \pm 0.003$.
In \cite{osc}, the stability of this fit was verified by a 
statistical bootstrap analysis
showing that the fits were not overly sensitive to any particular HII region.
In addition, the data from \cite{iz} was also included, yielding a \he4 mass
fraction \cite{osc}
\beq
Y_p = 0.234 \pm 0.003 \pm 0.005
\label{he4}
\eeq
The second uncertainty is an estimate of the systematic uncertainty in the
abundance determination. The solid box for \he4 in figure 1
represents the range (at 2$\sigma_{\rm stat} + \sigma_{\rm sys}$)
from (\ref{he4}). The dashed box extends this by $\sigma_{\rm sys}$.
With the addition of the newer data in \cite{iz},
the resulting \he4 abundance is also given by (\ref{he4}) with a smaller
statistical error of 0.002, although a case can also be made for a
somewhat lower primordial abundance of  $Y_p = 0.230 \pm .003$
by restricting to the most metal poor regions \cite{ost3}.  

The \li7 abundance
is also reasonably well known.
 In old,
hot, population-II stars, \li7 is found to have a very
nearly  uniform abundance. For
stars with a surface temperature $T > 5500$~K
and a metallicity less than about
1/20th solar (so that effects such as stellar convection may not be important),
the  abundances show little or no dispersion beyond that which is
consistent with the errors of individual measurements.
Indeed, as detailed in \cite{sp}, much of the work concerning
\li7 has to do with the presence or absence of dispersion and whether
or not there is in fact some tiny slope to a [Li] = $\log$ \li7/H + 12 vs.
T or [Li] vs. [Fe/H] relationship.
There is \li7 data from nearly 100 halo stars, from a 
 variety of sources. I will use the value given in \cite{mol} 
as the best estimate
for the mean \li7 abundance and its statistical uncertainty in halo stars 
\beq
{\rm Li/H = (1.6 \pm 0.1 {}^{+0.4}_{-0.3} {}^{+1.6}_{-0.5}) \times 10^{-10}}
\label{li}
\eeq
 The first error is statistical, and the second
is a systematic uncertainty that covers the range of abundances
derived by various methods. The solid box for \li7 in figure 1 represents
the 2$\sigma_{\rm stat} + \sigma_{\rm sys}$ range from (\ref{li}).
The third set of errors in Eq. (\ref{li}) accounts for
 the possibility that as much as half
of the primordial \li7 has been
destroyed in stars, and that as much as 30\% of the observed \li7 may have been
produced in cosmic ray collisions rather than in the Big Bang.
The dashed box in figure 1, accounts for this additional uncertainty.
 Observations of \li6,
Be, and B help constrain the degree to which these effects
play a role \cite{fossw}. For \li7, the uncertainties are clearly dominated by
systematic effects.

Turning to D/H, we have three basic types of abundance information:
1) ISM data, 2) solar system information, and perhaps 3) a primordial
abundance from quasar absorption systems.  The best measurement for ISM D/H
is \cite{linetal}
\beq
{\rm (D/H)_{ISM}} = 1.60\pm0.09{}^{+0.05}_{-0.10} \times 10^{-5}
\eeq
The lower bound from deuterium establishes an upper bound on $\eta$
which is robust and is shown by the lower right of the solid box
in figure 1.
 The solar abundance of D/H is inferred from two
distinct measurements of \he3. The solar wind measurements of \he3 as well as 
the low temperature components of step-wise heating measurements of \he3 in
meteorites yield the presolar (D + \he3)/H ratio, as D was 
efficiently burned to
\he3 in the Sun's pre-main-sequence phase.  These measurements 
indicate that \cite{scostv,geiss}
\beq
{\rm \left({D +~^3He \over H} \right)_\odot = (4.1 \pm 0.6 \pm 1.4) \times
10^{-5}}
\eeq
 The high temperature components in meteorites are believed to yield the true
solar \he3/H ratio of \cite{scostv,geiss}
\beq
{\rm \left({~^3He \over H} \right)_\odot = (1.5 \pm 0.2 \pm 0.3) \times
10^{-5}}
\label{he3}
\eeq
The difference between these two abundances reveals the presolar D/H ratio,
giving,
\beq
{\rm (D/H)_{\odot}} \approx (2.6 \pm 0.6 \pm 1.4) \times 10^{-5}
\eeq

It should be noted that recent measurements of surface abundances 
of HD on Jupiter
show a somewhat higher value for D/H,  D/H = $5 \pm 2 \times 10^{-5}$ 
\cite{nie}. If this value is confirmed and if fractionation
does not significantly alter the D/H ratio (as it was suspected to for 
previous measurements involving CH$_3$D), it may have an important 
impact on galactic chemical evolution models.  This value
is marginally consistent with the inferred meteoritic values. 

Finally, there have been several recent reported measurements of 
D/H is high redshift quasar absorption systems. Such measurements are in
principle capable of determining the primordial value for D/H and hence $\eta$,
because of the strong and monotonic dependence of D/H on $\eta$.
However, at present, detections of D/H  using quasar absorption systems
indicate both a 
high and  low value of D/H.  As such, it should be cautioned 
that these values may not
turn  out to represent the true primordial value and it is very unlikely 
that both are primordial and indicate an inhomogeneity \cite{cos2}.
The first of these measurements \cite{quas1} indicated a rather high D/H ratio,
D/H $\approx$ 1.9 -- 2.5 $\times 10^{-4}$.  A 
re-observation of the high D/H absorption system has been resolved into 
two components, both yielding high values with an average value of D/H = $1.9
^{+0.4}_{-0.3} \times 10^{-4}$ \cite{rh1}. Other 
high D/H ratios were reported in \cite{quas3}. However, there are reported low
values of D/H in other such systems  \cite{quas2} with values D/H $\simeq 2.5
\times 10^{-5}$, significantly lower than the ones quoted above. 
Though this primordial D/H value is consistent with the solar and present values
of D/H, it is not consistent (at the 2 $\sigma$ level) with the determinations
of D/H in Jupiter, if they are correct.
The range of quasar absorber D/H is shown by the dashed box in figure 1.
It is probably
premature to use either of these values as the primordial D/H abundance in 
an analysis of big
bang nucleosynthesis, but it is certainly encouraging that 
future observations may
soon yield a firm value for D/H. It is however important to 
note that there does
seem to be a  trend that over the history of the Galaxy, the D/H ratio  is
decreasing, something we expect from galactic chemical evolution.  
Of course the
total amount of deuterium astration that has occurred is still uncertain, and
model dependent.

There are also several types of \he3 measurements. As noted above, meteoritic
extractions yield a presolar value for \he3/H as given in Eq. (\ref{he3}).
In addition, there are several ISM measurements of \he3 in galactic HII
regions \cite{bbbrw} which show a wide dispersion which may be indicative 
of pollution or a bias \cite{orstv}
\beq
 {\rm \left({~^3He \over H} \right)_{HII}} \simeq 1 - 5 \times 10^{-5}
\eeq
There is also a recent ISM measurement of \he3 \cite{gg}
with
\beq
 {\rm \left({~^3He \over H} \right)_{ISM}} = 2.1^{+.9}_{-.8} \times 10^{-5}
\eeq
  Finally there are observations of \he3 in planetary
nebulae \cite{rood} which show a very high \he3 abundance of 
\he3/H $\sim 10^{-3}$.

Each of the light element isotopes can be made consistent with theory for a
specific range in $\eta$. Overall consistency of course requires that
the range in $\eta$ agree among all four light elements.
However, as will be argued below D and \he3 are far more sensitive to 
chemical evolution than \he4 or \li7 and as such a the direct comparison
between the theoretical predictions of the primordial abundances of
D and \he3 with the observational determination of their abundances is far more 
difficult.  Therefore in what follows I will restrict the comparison between
theory and observation to the two isotopes who suffer the least from the
effects of chemical evolution.

\bigskip

Monte Carlo techniques are proving to be a useful form of analysis for big
bang nucleosynthesis \cite{kr,kk,hata1}. In \cite{fo}, we performed just such an
analysis using only \he4 and \li7. It should be noted that in principle, two
elements should be sufficient for not only constraining the one parameter 
($\eta$) theory of BBN, but also for testing for consistency. 
We begin by establishing likelihood functions for the theory and
observations. For example, for \he4, the theoretical likelihood 
function takes the
form
\beq
L_{\rm BBN}(Y,Y_{\rm BBN}) 
  = e^{-\left(Y-Y_{\rm BBN}\left(\eta\right)\right)^2/2\sigma_1^2}
\label{gau}
\eeq
where $Y_{\rm BBN}(\eta)$ is the central value for the \he4 mass fraction
produced in the big bang as predicted by the theory at a given value of $\eta$,
and $\sigma_1$ is the uncertainty in that  value derived from the Monte Carlo
calculations \cite{hata1} and is a measure of the theoretical 
uncertainty in the
big bang calculation. Similarly one can write down an expression for the
observational likelihood function. Assuming a Gaussian to describe the
systematic uncertainty, the likelihood function for the observations would have 
take a form similar to that in (\ref{gau}).

A total likelihood 
function for each value of $\eta_{10}$ is derived by
convolving the theoretical
and observational distributions, which for \he4 is given by
\beq
{L^{^4{\rm He}}}_{\rm total}(\eta) = 
\int dY L_{\rm BBN}\left(Y,Y_{\rm BBN}\left(\eta\right)\right) 
L_{\rm O}(Y,Y_{\rm O})
\label{conv}
\eeq
An analogous calculation is performed \cite{fo} for \li7. 
The resulting likelihood
functions from the observed abundances given in Eqs. (\ref{he4}) 
  and (\ref{li})
is shown in Figure 2. As one can see 
there is very good agreement between \he4 and \li7 in the vicinity
of $\eta_{10} \simeq 1.8$.

\begin{figure}
\hspace{0.5truecm}
\epsfysize=9truecm\epsfbox{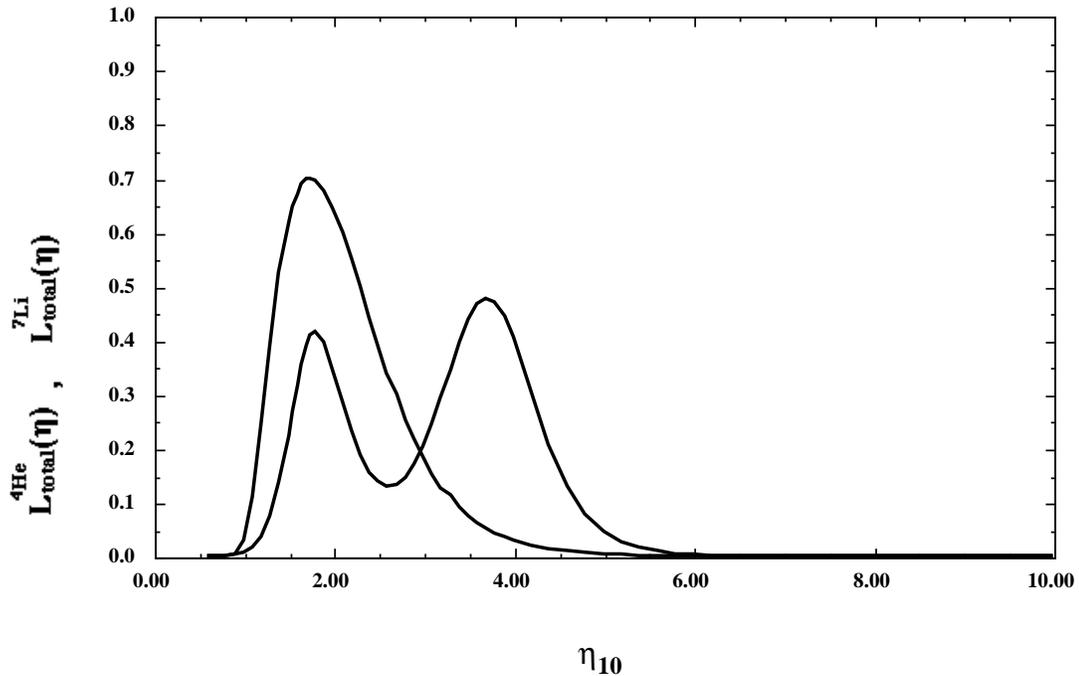}

\caption { \baselineskip=2ex Likelihood distribution for each of \he4 and \li7,
shown as a  function of $\eta$.  The one-peak structure of the \he4 curve
corresponds to its monotonic increase with $\eta$, while
the two-peaks for \li7 arise from its passing through a minimum.}
\label{fig:fig1}
\end{figure}

\begin{figure}
\hspace{0.5truecm}
\epsfysize=9truecm\epsfbox{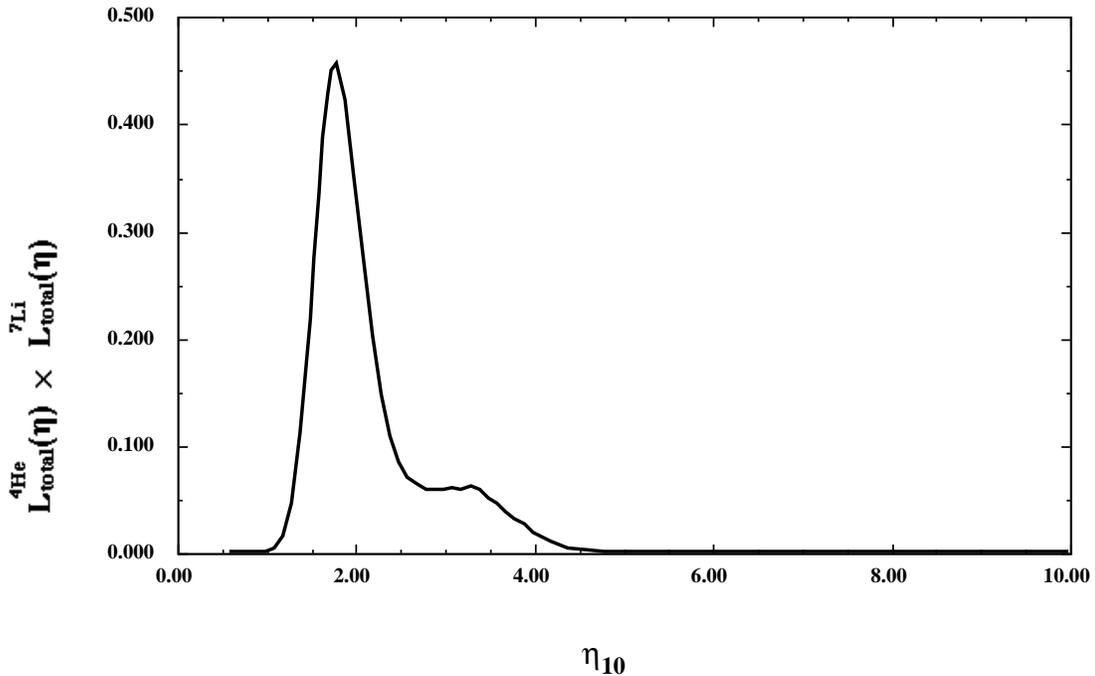}
\baselineskip=2ex
\caption { Combined likelihood for simultaneously fitting \he4 and \li7,
as a function of $\eta$.
}
\label{fig:fig2}
\end{figure}

\baselineskip=3ex

The combined likelihood, for fitting both elements simultaneously,
is given by the product of the two functions in Figure \ref{fig:fig1}
and is shown  in Figure \ref{fig:fig2}.
{}From Figure \ref{fig:fig1} it is clear that \he4 overlaps
the lower (in $\eta$) \li7 peak, and so one expects that 
there will be concordance
in an allowed range of $\eta$ given by the overlap region.  
This is what one finds in Figure \ref{fig:fig2}, which does
show concordance and gives a preferred value for $\eta$, 
$\eta_{10}  = 1.8^{+1}_{-.2}$ corresponding to 
\beq
\Omega_B h^2 = .006^{+.004}_{-.001}
\label{omega}
\eeq 

Thus,  we can conclude that 
the abundances of 
\he4 and \li7 are consistent, and select an $\eta_{10}$ range which
overlaps with (at the 95\% CL) the longstanding favorite
 range around $\eta_{10} = 3$.
Furthermore, by finding concordance  
using only \he4 and \li7, we deduce that
if there is problem with BBN, it must arise from 
D and \he3 and is thus tied to chemical evolution or the stellar evolution of
\he3. The most model-independent conclusion is that standard
BBN  with $N_\nu = 3$ is not in jeopardy, 
but there may be problems with our
detailed understanding of D and particularly \he3
chemical evolution. It is interesting to
note that the central (and strongly)  peaked
value of $\eta_{10}$ determined from the combined \he4 and\li7 likelihoods
is at $\eta_{10} = 1.8$.  The corresponding value of D/H is 1.8 $\times 
10^{-4}$, very close \cite{dar} to the high value  of D/H in quasar absorbers
\cite{quas1,rh1,quas3}.
Since  D and \he3 are monotonic functions of $\eta$, a prediction for 
$\eta$, based on \he4 and \li7, can be turned into a prediction for
D and \he3.  
 The corresponding 95\% CL ranges are D/H  $= (5.5 - 27)  \times
10^{-5}$ and  and \he3/H $= (1.4 - 2.7)  \times 10^{-5}$.

If we did have full confidence in the measured value of D/H in 
quasar absorption
systems, then we could perform the same statistical analysis 
using \he4, \li7, and
D. To include D/H, one would
proceed in much the same way as with the other two light elements.  We
compute likelihood functions for the BBN predictions as in
Eq. (\ref{gau}) and the likelihood function for the observations using
D/H = $(1.9 \pm 0.4) \times 10^{-4}$.  We are using only the high
 value of D/H
here. These are
then convolved as in Eq.  (\ref{conv}).  
In figure 4, the resulting normalized
distribution, $L^{{\rm D}}_{\rm total}(\eta)$ is super-imposed on
distributions appearing in figure 2. 
It is indeed startling how the three peaks, for
D, \he4 and \li7 are literally on top of each other.  In figure 5, 
the combined distribution is shown.
We now  have a very clean distribution and prediction 
for $\eta$, $\eta_{10}  = 1.75^{+.3}_{-.1}$ corresponding to $\Omega_B h^2 =
.006^{+.001}_{-.0004}$,
with the peak of the distribution at $\eta_{10} = 1.75$.  
The absence of any overlap with the high-$\eta$ peak of the \li7
distribution has considerably lowered the upper limit to $\eta$. 
Overall, the concordance limits in this case are dominated by the 
deuterium likelihood function.

\begin{figure}
\hspace{0.5truecm}
\epsfysize=9truecm\epsfbox{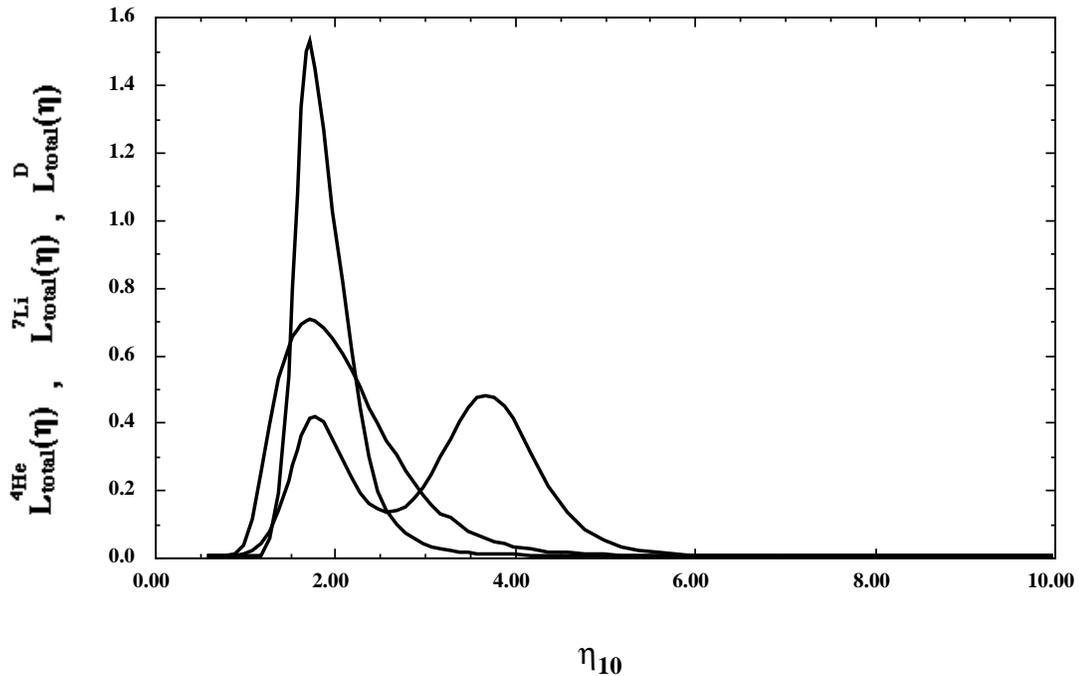}
\baselineskip=2ex
\caption { As in Figure 2, with the addition of the likelihood 
distribution for
D/H. }
\label{fig:fig4}
\end{figure}

\baselineskip=3ex

\begin{figure}
\hspace{0.5truecm}
\epsfysize=9truecm\epsfbox{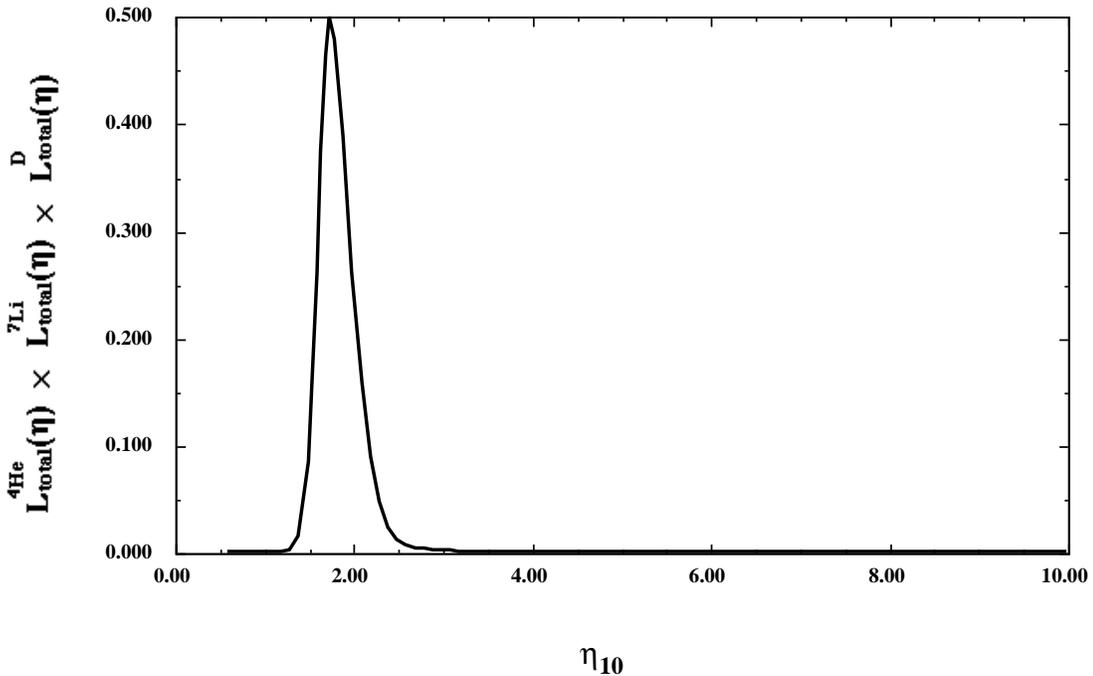}
\baselineskip=2ex
\caption { Combined likelihood for simultaneously fitting 
\he4 and \li7, and D 
as a function of $\eta$.
}
\label{fig:fig5}
\end{figure}

\baselineskip=3ex

This type of analysis also has the potential for placing constraints on
the degree of \li7 depletion in stars which is one of the major 
uncertainties associating the primordial abundance with the observations.
If depletion say by a factor of two were assumed rather than treated
as an uncertainty which has the effect of widening the distribution
functions in figures 2 and 4, the two lithium peaks would appear thinner
and split further apart \cite{fo}. As a result, there would be far less overlap
between the likelihood distributions of \he4 and \li7.  The combined
distribution would show two distinct peaks centered on $\eta_{10}$ = 1.3
and 5.0 with heights of 0.15 and 0.1 respectively with the same scaling
as shown in figures 3 and 5.  If in addition, we had confidence in the 
high redshift D/H measurements, the high value of D/H would eliminate
the high $\eta$ peak and leave a single blip at $\eta_{10} = 1.5$
with a relative height of 0.03 and
essentially exclude any depletion of \li7 in these stars.

For the most part I have concentrated on the high D/H measurements in the
likelihood analysis. If instead, we assume that the low value \cite{quas2}
of D/H = $(2.5 \pm 0.5) \times 10^{-5}$ is the primordial abundance,
then we can again compare the likelihood distributions as in figure 4,
now substituting the low D/H value. As one can see from figure 6, there 
is now hardly any overlap between the D and the \li7 distributions
and essentially no overlap with the \he4 distribution.  The combined distribution
shown in figure 7 is compared with that in figure 5.
Though one can not use this likelihood analysis to prove
 the correctness of the high
D/H measurements or the incorrectness of the low D/H measurements,
the analysis clearly shows the difference in compatibility between the
two values of D/H and the observational determinations of \he4 and \li7.
To {\em make} the low D/H measurement compatible, one would have to argue
for a shift upwards in \he4 to a primordial value of 0.249 (a shift by 0.015)
which is certainly not warranted by the data, and a \li7 depletion factor of 
about 3, which exceeds recent upper limits to the amount of depletion
\cite{cv}.

\begin{figure}
\hspace{0.5truecm}
\epsfysize=9truecm\epsfbox{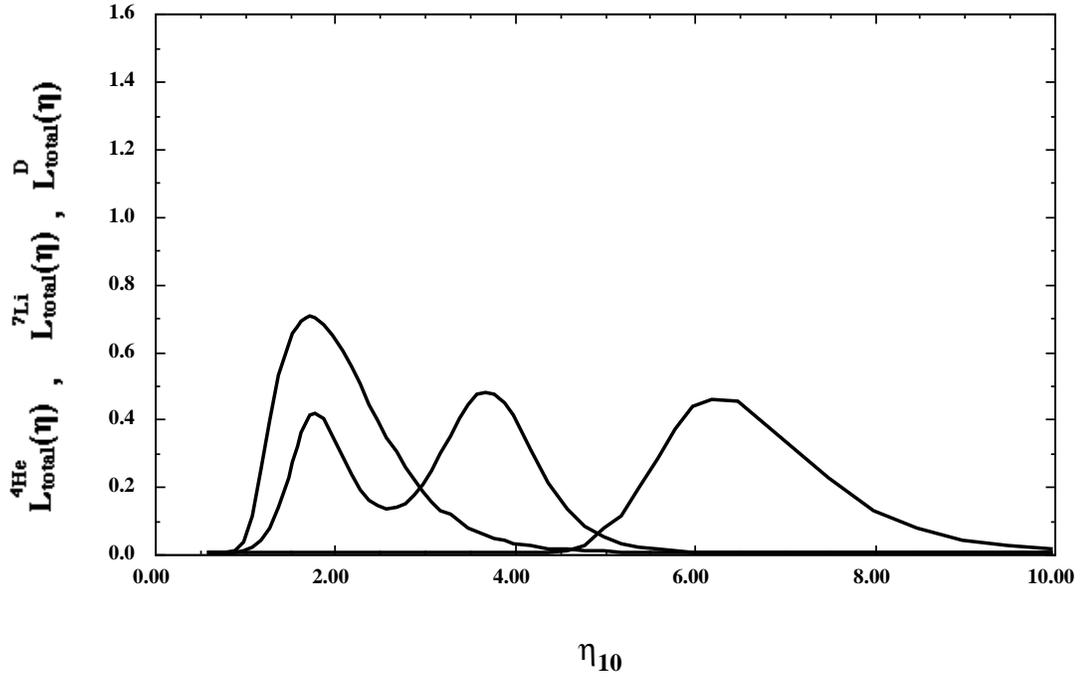}
\baselineskip=2ex
\caption { As in Figure 4, with the likelihood 
distribution for low 
D/H. }
\label{fig:fig6}
\end{figure}

\baselineskip=3ex

\begin{figure}
\hspace{0.5truecm}
\epsfysize=9truecm\epsfbox{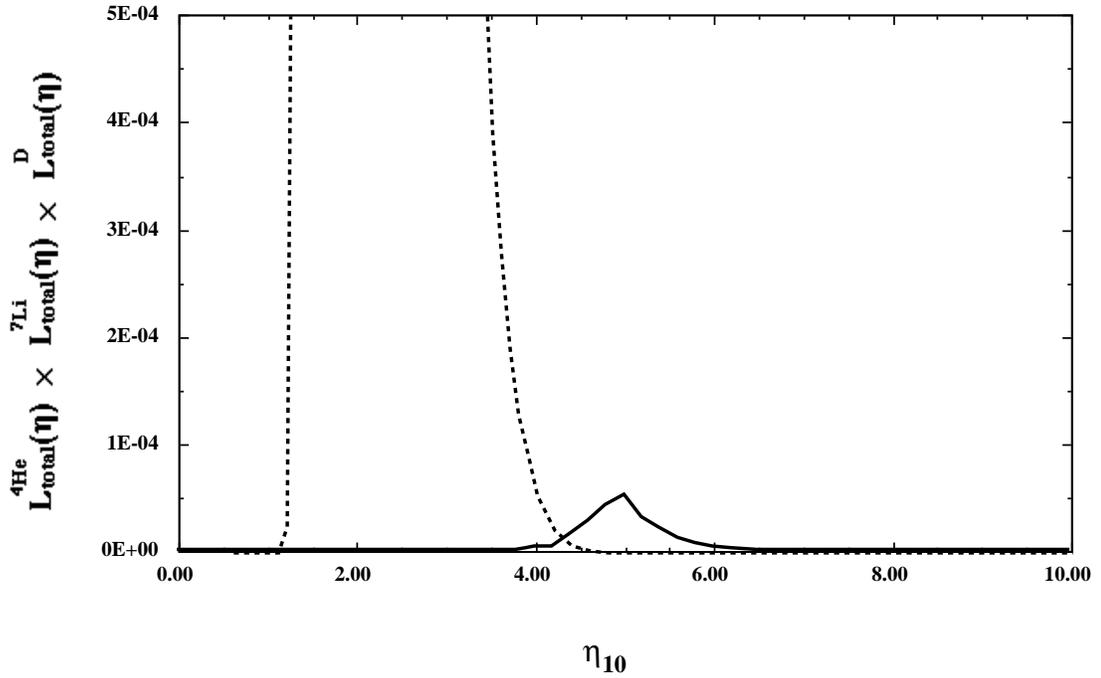}
\baselineskip=2ex
\caption { Combined likelihood for simultaneously fitting 
\he4 and \li7, and  low D/H
as a function of $\eta$. The dashed curve represents the combined distribution
shown in figure 5.
}
\label{fig:fig7}
\end{figure}

\baselineskip=3ex
\bigskip

The predictions by the above analysis for D and \he3 have important implications
for models of chemical evolution.
\he3 (together with D) has stood out in its importance for BBN, because 
it  provided a (relatively large) lower limit for the baryon-to-photon
ratio \cite{ytsso}, $\eta_{10} > 2.8$. This limit for a long 
time was seen to be
essential because it provided the only means for bounding $\eta$ from below
and in effect allows one to set an upper limit on the number of neutrino
flavors \cite{ssg}, $N_\nu$, as well as other constraints on particle physics
properties. That is, the upper bound to $N_\nu$ 
is strongly dependent on the lower bound to
$\eta$.  This is easy to see: for lower $\eta$, the \he4 abundance drops,
allowing for a larger $N_\nu$, which would raise the \he4 abundance.
However, for $\eta < 4 \times 10^{-11}$, corresponding to $\Omega h^2 \sim
.001-.002$, which is not too different from galactic mass densities, 
there is no
bound whatsoever on $N_\nu$ \cite{ossty}. Of course, with the improved data on
\li7, we do have lower bounds on $\eta$ which exceed $10^{-10}$.

 In \cite{ytsso}, it was argued that since stars (even massive stars) do not 
destroy \he3 in its entirety, we can obtain a bound on $\eta$ from an
upper bound to the solar D and \he3 abundances. One can in fact limit
\cite{ytsso,ped}
 the sum of primordial D and \he3 by applying the expression below at $t =
\odot$
\beq
{\rm \left({D + \he3 \over H} \right)_p \le \left({D \over H} \right)_t}
+ {1 \over g_3}{\rm  \left({\he3 \over H} \right)_t}
\label{he3lim}
\eeq
In (\ref{he3lim}), $g_3$ is the fraction of a star's initial D and \he3 which
survives as \he3. For $g_3 > 0.25$ as suggested by stellar models, 
and using the
solar data on D/H and
\he3/H, one finds $\eta_{10} > 2.8$. This limit on $\eta$
is shown by the upper left of the solid box in figure 1.
This argument has been improved
recently \cite{st} ultimately leading to a stronger limit \cite{hata2} 
$\eta_{10} > 3.8$ and a best estimate $\eta_{10} = 6.6 \pm 1.4$.
The stochastic approach used in Copi et al. \cite{cst} could
only lower the bound from 3.8 to about 3.5 when assuming as always that $g_3 >
0.25$.

The limit $\eta_{10} > 2.8$ derived using (\ref{he3lim}) is really a one
shot approximation.  Namely, it is assumed that material passes 
through a star no
more than once. To determine whether or not the solar (and present) 
values of D/H
and
\he3/H can be matched it is necessary to consider models of galactic chemical
evolution \cite{bt}. In the absence of stellar \he3 production, 
particularly by
low mass stars, it was shown \cite{vop} that there are indeed suitable choices
for a star formation rate and an initial mass function to: 1) match the D/H
evolution from a primordial value (D/H)$_p = 7.5 \times 10^{-5}$,
corresponding to $\eta_{10} = 3$, through the solar and ISM abundances, 
while 2)
at the same time keeping the \he3/H evolution relatively flat so as not to
overproduce \he3 at the solar and present epochs. This was achieved for $g_3 >
0.3$. Even for $g_3 \sim 0.7$, the present \he3/H could be matched, though the
solar value was found to be a factor of 2 too high. For (D/H)$_p 
\simeq 2 \times
10^{-4}$, corresponding to  $\eta_{10} \simeq 1.7$, models could be
found which destroy D sufficiently; however, overproduction of 
\he3 occurred unless
$g_3$ was tuned down to about 0.1 \cite{orstv}.

In the context of models of galactic chemical evolution, there is, however, 
 little justification a
priori for neglecting the production of \he3 in low mass
stars. Indeed, stellar models predict that considerable amounts of \he3 are
produced in stars between 1 and 3 M$_\odot$. For M $<$ 8M$_\odot$, Iben and
Truran \cite{it} calculate
\beq
(^3{\rm He/H})_f = 1.8 \times 10^{-4}\left({M_\odot \over M}\right)^2 
+ 0.7\left[({\rm D+~^3He)/H}\right]_i
\label{it}
\eeq
so that at $\eta_{10} = 3$, and ((D + \he3)/H)$_i = 9 \times 10^{-5}$,
$g_3(1 $M$_\odot$) = 2.7! It should be emphasized that this prediction is in
fact consistent with the observation of high \he3/H in planetary nebulae
\cite{rood}.

Generally, implementation of the \he3 yield in Eq. (\ref{it}) in chemical
evolution models leads to an overproduction of \he3/H particularly at the
solar epoch \cite{orstv,galli}. In Figure 8, the evolution of D/H and 
\he3/H is
shown as a function of time from \cite{scostv,orstv}. The solid curves 
show the
evolution in a  simple model of galactic chemical evolution with a star 
formation
rate proportional to the gas density and a power law IMF (see \cite{orstv}) for
details). The model was chosen to fit the observed deuterium 
abundances. However,
as one can plainly see, \he3 is grossly overproduced (the deuterium data is
represented by squares and \he3 by circles). Depending on the particular model
chosen,  it may be possible to come close to at least the upper 
end of the range of
the \he3/H observed in galactic HII regions \cite{bbbrw}, 
although the solar value
is missed by many standard deviations.

\begin{figure}
\hspace{4truecm}
\epsfysize=20truecm\epsfbox{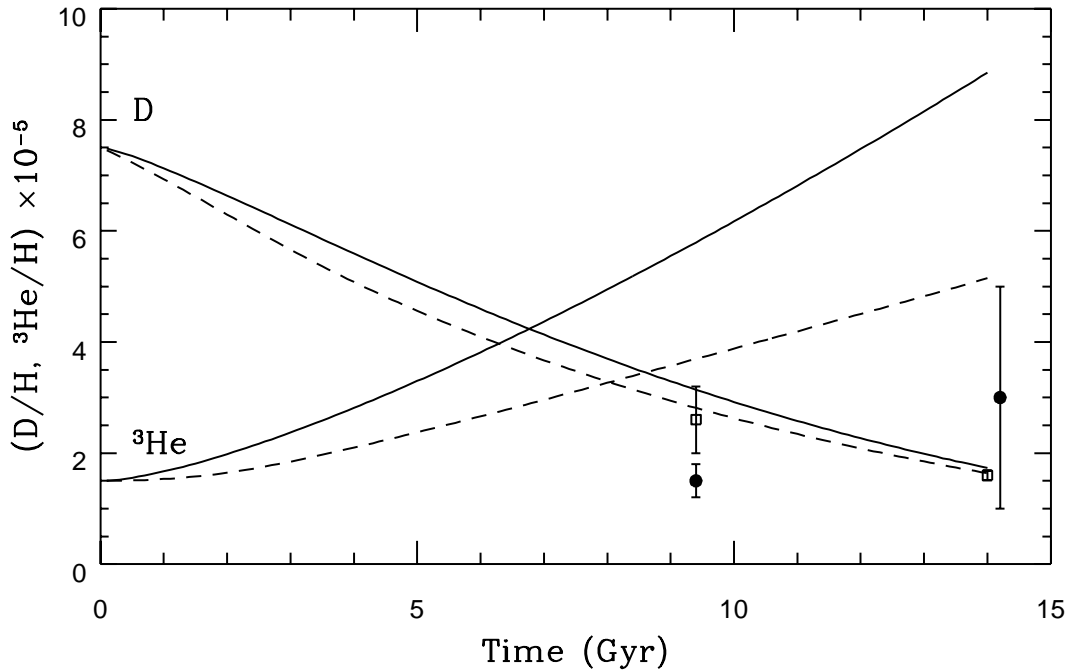}
\vspace{-10.5truecm}
\baselineskip=2ex
\caption { The evolution of D and \he3 with time.}
\end{figure}

\baselineskip=3ex
The overproduction of \he3 relative to the solar meteoritic value seems to be a
generic feature of chemical evolution models when \he3 production in low mass
stars is included. In \cite{scostv}, a more extreme model of galactic chemical
evolution was tested.  There, it was assumed that the initial mass function
was time dependent in such a way so as to favor massive stars early on (during
the first two Gyr of the galaxy).  Massive stars are preferential from the
point of view of destroying \he3.  However, massive stars are also proficient
producers of heavy elements and in order to keep the metallicity of the disk
down to acceptable levels, supernovae driven outflow was also included.
The degree of outflow was limited roughly by the observed metallicity in the
intergalactic gas in clusters of galaxies. One further assumption was 
necessary;
we allowed the massive stars to lose their \he3 depleted hydrogen
envelopes prior
to explosion.  Thus only the heavier elements were expulsed from the galaxy.
With all of these (semi-defensible) assumptions, \he3 was still overproduced
at the solar epoch, as shown by the dashed curve in Figure 8. Though there
certainly is an improvement in the evolution of \he3 without reducing the
yields of low mass stars, it is hard to envision much further reduction in
the solar \he3 predicted by these models. 

The model which results in an evolution given by figure 8,
has a modest amount of deuterium destruction.  If $\eta_{10}$ is close to 1.8
as predicted by \he4 and \li7 or as may be inferred from the 
high D/H QSO absorber measurements, the primordial value of D/H is 
higher than that depicted in figure 8, and requires substantially
more destruction of D.  In Scully et al. \cite{scov}, a dynamically generated
supernovae wind model was coupled to models of galactic chemical evolution
with the aim of reducing a primordial D/H abundance of 2 $\times 10^{-4}$
to the present ISM value without overproducing heavy elements and 
remaining within the other observational constraints typically 
imposed on such models.  \he3 remains a challenge to models of 
chemical and stellar evolution.

\bigskip

As indicated earlier, 
the presence of a lower bound on $\eta$ allows us to place an upper
bound to the number of neutrino flavors.  From (\ref{he3lim}), the limit
$\eta_{10} > 2.8$ corresponds to the limit $N_\nu < 3.3$ \cite{wssok}.
However, it should be noted that for values of $\eta$ larger than 2.8,
the central or best-fit value for $N_\nu$ is closer to 2 \cite{osa,kk,hata2}
and the upper bound is actually found to be 
much smaller with a careful treatment of the
uncertainties, $N_\nu \la 3.1$ \cite{osa,kk}, though this limit
is relaxed somewhat when the distribution for $N_\nu$ is renormalized \cite{osb}.
The range in $\eta$ of 6.6 $\pm 1.4$, corresponds to an even tighter
limit on $N_\nu = 2.0 \pm 0.3$ \cite{hata2} and indicates a problem
when trying to make use of D and \he3 in conjunction with \he4.

Given the magnitude of the problems concerning \he3, it would seem unwise to
make any strong conclusion regarding big bang nucleosynthesis 
which is based on \he3.  Perhaps as well some caution is 
deserved with regard to the recent D/H
measurements, although if the present trend continues and is 
verified in several
different quasar absorption systems, then D/H will certainly become our best
measure for the baryon-to-photon ratio. Just as \he4 and \li7 were sufficient to
determine a value for $\eta$, in so doing, a limit on $N_\nu$ can be obtained
as well \cite{fo,oth3}. The resulting best-fit to $N_\nu$ based on \he4 and
\li7 was found to be \cite{fo}
\beq
N_\nu = 3.0 \pm 0.2 \pm 0.4^{+ 0.1}_{- 0.6} 
\label{Nlim2}
\eeq
thus preferring the 
standard model result of $N_\nu = 3$ and leading to $N_\nu < 3.90$ 
at the 95 \% CL level when adding the errors in quadrature.
In (\ref{Nlim2}), the first set of errors are the
statistical uncertainties primarily from the observational
determination of $Y$ and the measured error in the neutrino half life
$\tau_n$. The second set of errors is the systematic uncertainty
arising solely from $^4$He, and the last set of errors from the
uncertainty in the value of $\eta$ and is determined by the combined
likelihood functions of \he4 and \li7, ie taken from
Eq.\ (\ref{omega}). A similar result is obtained when D/H is included in the
analysis.

The implications of the resulting predictions from big bang nucleosynthesis
on dark matter are clear.  First, if $\Omega = 1$ (as predicted by 
inflation), and $\Omega_B \la 0.1$ which is certainly a robust
conclusion based on D/H, then non baryonic dark matter is a necessity.
Second, on the scale of small groups of galaxies, $\Omega \ga 0.05$,
and is expected to sample the dark matter in galactic halos.
This value is probably larger than the best estimate for $\Omega_B$
from equation (\ref{omega}). $\Omega_B h^2 = 0.0065$ corresponds to 
$\Omega_B = 0.025$ for $h = 1/2$.  In this event,
some non-baryonic dark matter in galactic halos is required. 
This conclusion is unchanged
by the inclusion of the high D/H measurements in QSO absorbers.
In contrast \cite{2d}, the low D/H measurements would imply that
$\Omega_B h^2 = 0.023$ allowing for the possibility that $\Omega_B
\simeq 0.1$.  In this case, no non-baryonic dark matter is
required in galactic halos.  However, I remind the reader that
the low D/H is at present not consistent with either the observations of
\he4 nor \li7 and their interpretations as being primordial abundances.

To summarize on the subject of big bang nucleosynthesis, 
I would assert that one
can conclude that the present data on the abundances of the light element
isotopes are consistent with the standard model of big bang 
nucleosynthesis. Using
the the isotopes with the best data, \he4 and
\li7, it is possible to constrain the theory and obtain a best value for the
baryon-to-photon ratio of $\eta_{10} = 1.8$, a corresponding 
value $\Omega_B h^2 =
.0065$ and
\beq
\begin{array}{ccccc}
1.4 & < & \eta_{10} & < &  3.8  \qquad 95\% {\rm CL} \nonumber \\
.005 & < & \Omega_B h^2 & < & .014  \qquad 95\% {\rm CL}
\label{res2}
\end{array}
\eeq
For $0.4 < h < 1$, we have a range $ .005 < \Omega_B < .09$.
This is a rather low value for the baryon density
 and would suggest that much of the galactic dark matter is
non-baryonic \cite{vc}. These predictions are in addition 
consistent with recent
observations of D/H in quasar absorption systems which show a high value.
Difficulty remains however, in matching the solar \he3 abundance, suggesting a
problem with our current understanding of galactic chemical evolution or the
stellar evolution of low mass stars as they pertain to \he3.

\vskip 1in
\vbox{
\noindent{ {\bf Acknowledgments} } \\
\noindent    I would like to thank M. Cass\'{e}, C. Copi,
B. Fields, K. Kainulainen, R. Rood,
D. Schramm, S. Scully,
G. Steigman, D. Thomas, J. Truran, T. Walker, 
E. Vanginoni-Flam for many enjoyable collaborations. 
This work was supported in part by DOE grant
DE--FG02--94ER--40823.}



\baselineskip=2ex
\begin{thebibliography}{99}

\bibitem{wssok} T.P. Walker, G. Steigman, D.N. Schramm, 
 K.A. Olive and K. Kang, {\it Ap.J.} {\bf 376}
 (1991) 51.

\bibitem{fo} B.D. Fields and K.A. Olive, {\it Phys. Lett.} {\bf B368}
(1996) 103; B.D. Fields, K. Kainulainen, D. Thomas, and K.A. Olive,
astro-ph/9603009, {\it New Astronomy} {\bf 1} (1996) 77.

\bibitem{rpp} Review of Particle Properties, {\it Phys. Rev.} {\bf 54} (1996) 1.

\bibitem{hata1} N. Hata, R.J. Scherrer, G. Steigman, D. Thomas, and T.P.
Walker, {\it Ap.J.} {\bf 458} (1996) 637.

\bibitem{p} B.E.J. Pagel, E.A. Simonson, R.J. Terlevich and M. Edmunds, 
{\it MNRAS} {\bf 255} (1992) 325.

\bibitem{evan} E. Skillman et al., {\it Ap.J. Lett.} (in preparation) 1996.

\bibitem{iz} Y.I. Izatov, T.X. Thuan, and V.A. Lipovetsky,
{\it Ap.J.} {\bf 435} (1994) 647; preprint 1996.

\bibitem{osa} K.A. Olive and G. Steigman,  {\it Ap.J. Supp.}
 {\bf 97} (1995) 49.

\bibitem{osc} K.A. Olive and S.T. Scully, {\it IJMPA} {\bf 11} (1995) 409.

\bibitem{ost3} K.A. Olive and G. Steigman, in preparation.

\bibitem{sp} M. Spite, P. Francois, P.E. Nissen, and F. Spite, {\it A.A.} {\bf
307} (1996) 172; F. Spite,  to be published in the Proceedings of the  IInd
Rencontres du Vietnam: The Sun and Beyond, ed. J. Tran Thanh Van, 1996.
 
\bibitem{mol} P. Molaro, F. Primas, and P. Bonifacio, {\it A.A.} {\bf 295 }
(1995) L47.

\bibitem{fossw} T.P. Walker, G. Steigman, D.N. Schramm, K.A. Olive
and B. Fields, {\it Ap.J.} {\bf 413} (1993) 562;  K.A. Olive, and D.N.
 Schramm,
{\it Nature} {\bf 360} (1993) 439; G. Steigman, B. Fields, K.A. Olive,
D.N. Schramm, and  T.P.  Walker, Ap.J. {\bf 415} (1993) L35.

\bibitem{linetal} J.L. Linsky, et al., {\it Ap.J.} {\bf 402} (1993) 694;
J.L. Linsky, et al.,{\it Ap.J.} {\bf 451} (1995) 335.

\bibitem{scostv} S.T. Scully, M. Cass\'{e}, K.A. Olive, D.N. Schramm,
J.W. Truran, and E. Vangioni-Flam, astro-ph/0508086, {\it Ap.J.} {\bf 462} 
(1996)
960.

\bibitem{geiss} J. Geiss,  in {\it Origin
 and Evolution of the Elements}, eds.\ N.\ Prantzos,
E.\ Vangioni-Flam, and M.\ Cass\'{e}
(Cambridge: Cambridge University Press, 1993), p.~89.

\bibitem{nie} H.B. Niemann, et al. {\it Science}  {\bf 272} (1996) 846.

\bibitem{cos2} C. Copi, K.A. Olive, and D.N. Schramm, astro-ph/9606156 (1996).

\bibitem{quas1} R.F. Carswell, M. Rauch, R.J. Weymann, A.J. Cooke, and
J.K. Webb, {\it MNRAS} {\bf 268} (1994) L1; A. Songaila, L.L. Cowie,  
C. Hogan, and M. Rugers, {\it Nature} {\bf 368} (1994) 599.

\bibitem{rh1} M. Rugers and C.J. Hogan, {\it Ap.J.} {\bf 459} (1996)  L1.

\bibitem{quas3}  M. Rugers and C.J. Hogan, {\it A.J.} {\bf 111} (1996) 2135;
 R.F. Carswell, et al. {\it MNRAS} {\bf 278} (1996) 518;
E.J. Wampler, et al.  astro-ph/9512084, {\it A.A.} (1996) in press.

\bibitem{quas2} D. Tytler, X.-M. Fan, and S. Burles, {\it Nature} {\bf 381}
 (1996) 207; S. Burles and D. Tytler, {\it Ap.J.} {\bf 460} (1996) 584.

\bibitem{bbbrw} D.S. Balser, T.M. Bania, C.J. Brockway, R.T.
Rood, and T.L. Wilson, {\it Ap.J.} {\bf 430} (1994) 667.

\bibitem{orstv} K.A. Olive, R.T. Rood, D.N. Schramm, J.W. Truran, and E.
Vangioni-Flam, {\it Ap.J.} {\bf 444} (1995) 680.

\bibitem{gg} G. Gloeckler, and J. Geiss,  {\it Nature} {\bf 381} (1996) 210.

\bibitem{rood} R.T. Rood, T.M. Bania, and T.L. Wilson,  Nature 
{\bf 355} (1992)
618; R.T. Rood, T.M. Bania, T.L. Wilson, and D.S. Balser, 1995,
in {\it
 the Light Element Abundances, Proceedings of the ESO/EIPC Workshop},
ed. P. Crane, (Berlin:Springer), p. 201.

\bibitem{kr} L.M. Krauss and P. Romanelli,  {\it Ap.J.} {\bf 358} (1990) 47;
L.M. Krauss and P.J. Kernan, {\it Phys. Lett.} {\bf B347} (1995) 347;
M. Smith, L. Kawano, and R.A. Malaney, 
{\it Ap.J. Supp.} {\bf 85} (1993) 219.

\bibitem{kk} P.J. Kernan and L.M. Krauss, {\it Phys. Rev. Lett.} 
{\bf 72} (1994) 3309.

\bibitem{dar} A. Dar, {\it Ap.J.} {\bf 449} (1995) 550.

\bibitem{cv} S. Vauclair and C. Charbonnel, {\it A.A.} {\bf 295} (1995) 715.

\bibitem{ytsso} J. Yang, M.S. Turner, G. Steigman, D.N. Schramm, and
 K.A. Olive, {\it Ap.J.} {\bf 281} (1984) 493.

\bibitem{ssg} G. Steigman, D.N. Schramm, and J. Gunn,
  {\it Phys. Lett.} {\bf B66} (1977) 202.

\bibitem{ossty}   K.A. Olive, D.N. Schramm, G. Steigman, M.S. Turner, and J.
Yang,  {\it Ap.J.} {\bf 246} (1981) 557.

\bibitem{ped} D. Black, {\it Nature Physical Sci.}, {\bf 24} (1971) 148.

\bibitem{st} G. Steigman and M. Tosi, Ap.J. {\bf 401} (1992) 15;
G. Steigman and M. Tosi, Ap.J. {\bf 453} (1995) 173.

\bibitem{hata2} N. Hata, R. J. Scherrer, G. Steigman, D. Thomas,
T. P. Walker, S. Bludman and P. Langacker, Phys. Rev. Lett. {\bf 75} (1995) 3977.

\bibitem{cst} C.J. Copi, D.N. Schramm, and M.S. Turner,  Ap.J. {\bf 455} (1995)
L95.

\bibitem{bt} B.M. Tinsley, {\it  Fund Cosm Phys} {\bf 5} (1980) 287.

\bibitem{vop} E. Vangioni-Flam, K.A. Olive, and N. Prantzos, {\it Ap.J.} {\bf 
427} (1994) 618.

\bibitem{it} I. Iben, and J.W. Truran, {\it Ap.J.} {\bf 220} (1978) 980.

\bibitem{galli} D. Galli, F. Palla, F. Ferrini, and U. Penco,
Ap.J. {\bf 443} (1995) 536; D. Dearborn, G. Steigman, and M. Tosi, {\it Ap.J.}
{\bf 465} (1996) in press.

\bibitem{scov} S. Scully, M. Cass\'{e}, K.A. Olive, and 
E. Vagioni-Flam, astro-ph/9607106, {\it Ap.J.} (1996) in press.

\bibitem{osb} K.A. Olive and G. Steigman, {\it Phys. Lett.} 
{\bf B354 } (1995) 357.

\bibitem{oth3} K.A. Olive and D. Thomas, in preparation.

\bibitem{2d} C.Y. Cardall and G.M. Fuller, astro-ph/9603071 (1996);
N. Hata, G. Stiegman, S. Bludman, and P. Langacker, 
astro-ph/9603087 (1996).

\bibitem{vc} E. Vangioni-Flam and M. Cass\'{e}, Ap.J. {\bf 441} (1995) 471.


\end{thebibliography}
\end{document}